\documentclass[runningheads]{svmult}

\usepackage{makeidx}   
\usepackage{graphicx}  
\usepackage{subeqnar}  
\usepackage{multicol}  
\usepackage{physprbb}  
\makeindex             

\newcommand\lsim{\mathrel{\rlap{\lower 4pt \hbox{\hskip 1pt
$\sim$}}\raise 1pt \hbox {$<$}}}
\newcommand\gsim{\mathrel{\rlap{\lower 4pt \hbox{\hskip 1pt
$\sim$}}\raise 1pt \hbox {$>$}}}

\newcommand{\msun}{$M_{\odot}$}
\newcommand\lam{$\lambda$}
\newcommand\mdot{$\dot M$}
\newcommand\msy{$M_\odot$ yr$^{-1}$}
\newcommand\kms{km s$^{-1}$}

\begin{document}
\title*{Type Ia Supernovae: Progenitors and Diversities}
\toctitle{Type Ia Supernovae: Progenitors and Diversities}
%
%
\titlerunning{Type Ia Supernovae: Progenitors and Diversities}
%
\author{Ken'ichi Nomoto\inst{1}
\and Tatsuhiro Uenishi\inst{1}
\and Chiaki Kobayashi\inst{2}
\and Hideyuki Umeda\inst{1}
\and Takuya Ohkubo\inst{1}
\and Izumi Hachisu\inst{3}
\and Mariko Kato\inst{4}}
\authorrunning{Ken'ichi Nomoto et al. }
%
%

\institute{Department of Astronomy, School of Science, University of Tokyo
\and Max-Planck-Institut f\"ur Astrophysik, Garching
\and Department of Earth Science \& Astronomy, University of Tokyo
\and Department of Astronomy, Keio University}
\maketitle              

\vspace*{-70mm}
\noindent
{\scriptsize Invited Review published in "From Twilight to Highlight: The Physics of Supernovae," eds. W. Hillebrandt \& B. Leibundgut, 
ESO/Springer Series "ESO Astrophysics Symposia" (Berlin: Springer) p.115--127(2003).}
\vspace*{60mm}

\begin{abstract}

A key question for supernova cosmology is whether the peak luminosities
of Type Ia supernovae (SNe Ia) are sufficiently free from the effects of
cosmic and galactic evolution.  To answer this question, we review the
currently popular scenario of SN Ia progenitors, i.e., the single
degenerate scenario for the Chandrasekhar mass white dwarf (WD) models.
We identify the progenitor's evolution with two channels: (1) the WD+RG
(red-giant) and (2) the WD+MS (near main-sequence He-rich star)
channels.  The strong wind from accreting WDs plays a key role, which
yields important age and metallicity effects on the evolution.  

We suggest that the variation of the carbon mass fraction $X$(C) in
the C+O WD (or the variation of the initial WD mass) causes the
diversity of SN Ia brightness.  This model can explain the observed
dependence of SNe Ia brightness on the galaxy types.  We then predict
how SN Ia brightness evolves along the redshift (with changing
metallicity and age) for elliptical and spiral galaxies.  Such
evolutionary effects along the redshift can be corrected as has been
made for local SNe Ia.

We also touch on several related issues: (1) the abundance pattern of
stars in dwarf spheroidal galaxies in relation to the metallicity effect
on SNe Ia, (2) effects of angular momentum brought into the WD in
relation to the diversities and the fate of double degenerates, and (3)
possible presence of helium in the peculiar SN Ia 2000cx in relation to
the sub-Chandrasekhar mass model.

\end{abstract}

\section{Introduction}

Relatively uniform light curves and spectral evolution of Type Ia
supernovae (SNe Ia) have led to the use of SNe Ia as a ``standard
candle'' to determine cosmological parameters.  Whether a statistically
significant value of the cosmological constant can be obtained depends
on whether the peak luminosities of SNe Ia are sufficiently free from
the effects of cosmic and galactic evolutions \cite{leib01}.

SNe Ia have been widely believed to be a thermonuclear explosion of a
mass-accreting white dwarf (WD).  However, the immediate progenitor
binary systems have not been clearly identified yet \cite{bra95}.  In
order to address the above questions regarding the nature of
high-redshift SNe Ia, we need to identify the progenitors systems and
examine the ``evolutionary'' effects (or environmental effects) on
those systems \cite{ume99b}.

Here we review several issues such as Chandra vs. sub-Chandra, double
degenerates vs. single degenerate (AIC vs. SN Ia), C/O ratio, and
rotation.  For complete discussion, see earlier reviews on SN Ia
progenitors \cite{arn96,bra98,hil00,liv00,nom94,nom97,nom00}.

\section{Evolution of progenitor systems}

\subsubsection{Chandra vs. Sub-Chandra:}

There exist two models proposed as progenitors of SNe Ia: 1) the
Chandrasekhar mass model, in which a mass-accreting carbon-oxygen (C+O)
WD grows its mass $M_{\rm WD}$ up to the critical mass $M_{\rm Ia}
\simeq 1.37-1.38 M_\odot$ near the Chandrasekhar mass and explodes as an
SN Ia (e.g., \cite{nom84,nom94}), and 2) the sub-Chandrasekhar mass
model, in which an accreted layer of helium atop a C+O WD ignites
off-center for a WD mass well below the Chandrasekhar mass (e.g.,
\cite{arn96}). The early time spectra of the majority of SNe Ia are in
excellent agreement with the synthetic spectra of the Chandrasekhar mass
models, while the spectra of the sub-Chandrasekhar mass models are too
blue to be comparable with the observations \cite{hof96,nug97}.
However, the peculiar SN Ia 2000cx might be the sub-Chandrasekhar mass
explosion as will be discussed in \S6.

\begin{figure}[t]
	\begin{center}
	\begin{minipage}[t]{0.8\textwidth}
		\includegraphics[width=0.85\textwidth]{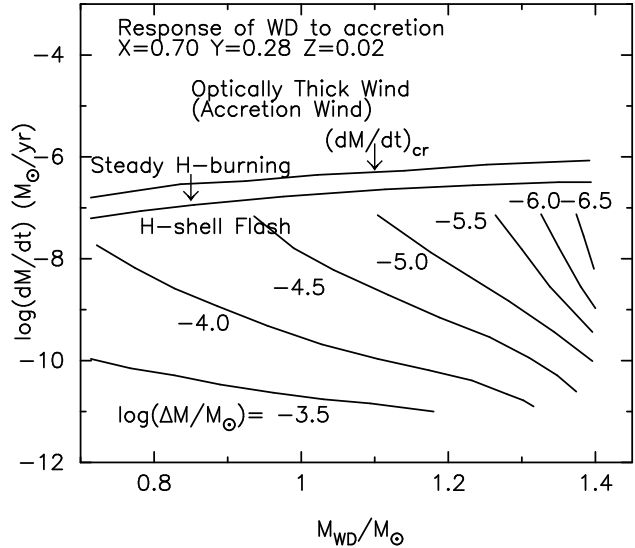}
	\end{minipage}
	\end{center}
\caption{
The nature of hydrogen burning on accreting WD as functions of the WD
mass and the accretion rate \cite{nom82a}.  Above the steady burning
regime, the WD blows optically thick wind \cite{hac01}.
\label{nomotoF1}}
\end{figure}

\subsubsection {Double Degenerates vs. Single Degenerate: }

     For the evolution of accreting WDs toward the Chandrasekhar mass,
two scenarios have been proposed: 1) a double degenerate (DD)
scenario, i.e., merging of double C+O WDs \cite{ibe84,web84}, and 2) a
single degenerate (SD) scenario, i.e., accretion of hydrogen-rich
matter via mass transfer from a binary companion (e.g.,
\cite{nom82a,nom94}).  The issue of DD vs. SD is still debated,
although theoretical modeling has indicated that the merging of WDs
leads to the accretion-induced collapse (AIC) rather than SN Ia explosion
\cite{saio98,seg97}.  Whether the effect of rotation on accretion 
changes the conclusion on AIC vs. SN Ia will be discussed in \S5.2.

\subsection {White dwarf winds}

     In the SD Chandrasekhar mass model, a WD explodes as a
SN Ia only when its rate of the mass accretion ($\dot M$) is in a
certain narrow range (Fig. \ref{nomotoF1} \cite{nom82a,hac01}).
In particular, if $\dot M$ exceeds the critical rate $\dot M_{\rm b}$,
the accreted matter extends to form a common envelope \cite{nom79}.
Here $\dot M_{\rm b}$ is
the rate at which steady burning can process the accreted hydrogen
into helium as 
$\dot M_{\rm b} \approx 0.75 \times10^{-6} \left({M_{\rm WD} \over {M_\odot}}
 - 0.40\right) M_\odot {\rm ~yr}^{-1}$.

     For \mdot $\gsim$ $\dot M_{\rm b}$, the strong peak of Fe opacity
\cite{igl93} drives the radiation-driven wind from the WD
\cite{hac99b}.  If the wind is sufficiently strong, the WD can
avoid the formation of a common envelope and steady hydrogen burning
increases its mass continuously at a rate $\dot M_{\rm b}$ by blowing
the extra mass away in a wind.  For \mdot $\lsim$ 0.5 $\dot M_{\rm b}$,
hydrogen shell burning becomes unstable to trigger weak shell flashes
but still burns a large fraction of accreted hydrogen.  Recurrent novae
appear in the upper-right region of Fig. \ref{nomotoF1}.  In this way,
strong winds from the accreting WD play a key role to increase the WD
mass to $M_{\rm Ia}$.

\subsection {Progenitor binary systems}

For the actual binary systems which grow $M_{\rm WD}$ to $M_{\rm Ia}$,
the following two systems are appropriate.

\subsubsection{WD+RG system (Symbiotic system):} 

This is a symbiotic binary system consisting of a WD and a low mass
red-giant (RG) \cite{hac99b}.  The immediate progenitor binaries in this
symbiotic channel to SNe Ia may be observed as symbiotic stars, luminous
supersoft X-ray sources, or recurrent novae like T CrB or RS Oph,
depending on the wind status \cite{hac01}.

\subsubsection{WD+MS system (Super-soft system):} 

This is a system consisting of a mass-accreting WD and a lobe-filling,
more massive, slightly evolved main-sequence or sub-giant star.
In this scenario, a C+O WD is originated, not from an AGB star with a
C+O core, but from a red-giant star with a helium core of $\sim
0.8-2.0 M_\odot$.  The helium star, which is formed after the first
common envelope evolution, evolves to form a C+O WD of $\sim 0.8-1.1
M_\odot$ with transferring a part of the helium envelope onto the
secondary main-sequence star \cite{hac99a}.

A part of the progenitor systems are identified as the luminous
supersoft X-ray sources \cite{heu92} during steady H-burning (but
without wind to avoid extinction), or the recurrent novae like U Sco
if H-burning is weakly unstable \cite{hac01}.  Actually these objects
are characterized by the accretion of helium-rich matter.

\subsubsection{Realization frequency:}

     The rate of SNe Ia originating from these channels in our Galaxy is
estimated with equation (1) of \cite{ibe84}.  The realization
frequencies of SNe Ia through the WD+RG and WD+MS channels are estimated
as $\sim$ 0.0017 yr$^{-1}$ (WD+RG) and $\sim$ 0.001 yr$^{-1}$ (WD+MS),
respectively.  The total SN Ia rate of the WD+MS/WD+RG systems becomes
$\sim$ 0.003 yr$^{-1}$, which is close enough to the inferred rate of
our Galaxy.

\subsection{Metallicity dependence of type Ia supernovae}

The optically thick winds are driven by a strong peak of OPAL opacity
due to iron lines.  Thus the wind velocity $v_{\rm w}$ is lower for
lower [Fe/H].  The SN Ia regions are much smaller for lower metallicity,
and very few SN Ia occurs at [Fe/H] $\le -1.1$ in this model.
It is possible to test such metallicity effects on SNe Ia with the
chemical evolution of galaxies.

In the one-zone uniform model for the chemical evolution of the solar
neighborhood, the heavy elements in the metal-poor stars originate
from the mixture of the SN II ejecta of various progenitor masses.
The abundances averaged over the progenitor masses of SNe II predicts
[O/Fe] $\sim 0.45$ (e.g., \cite{tnh96}).  Later SNe Ia start ejecting
mostly Fe, so that [O/Fe] decreases to $\sim 0$ around [Fe/H] $\sim
0$.  The low-metallicity inhibition of SNe Ia predicts that the
decrease in [O/Fe] starts at [Fe/H] $\sim -1$.  Such an evolution of
[O/Fe] well explains the observations \cite{kob98,kob00}.

We should note that some anomalous stars have [O/Fe] $\sim$ 0 at
[Fe/H] $\lsim -1$.  The presence of such stars, however, is not in
conflict with the metallicity dependence of SNe Ia, but can be
understood as follows: The formation of such anomalous stars (and the
diversity of [O/Fe] in general) indicates that the interstellar
materials were not uniformly mixed but contaminated by only a few SNe
II (or even single SN II) ejecta.  The Fe and O abundances produced by
a single SN II vary depending mainly on the mass of the progenitor.
Relatively smaller mass SNe II ($13-15 M_\odot$) and higher explosion
energies tend to produce [O/Fe] $\sim 0$ \cite{tnh96,ume02}.  Those
metal-poor stars with [O/Fe] $\sim 0$ may be born from the
interstellar medium polluted by such SNe II.  Alternatively, such
stars were born in nearby dwarf spheroidal galaxies and captured by
our Galaxy (see below).

\subsection{Abundances in dwarf spheroidal galaxies}

The chemical abundances of individual stars in local dwarf spheroidal
galaxies (dSph) have been measured (e.g., \cite{shet02} for recent
observations and references therein).  These stars have [Fe/H] $< -1$
but the ratios between the $\alpha$ elements and Fe are found to be as
low as [$\alpha$/Fe] $\sim$ 0 being significantly lower than the
metal-poor halo-stars in our Galaxy.  If the low [$\alpha$/Fe] was due
to the Fe-enrichment by SNe Ia, it implies that SNe Ia appeared in low
metallicity environment, which may be difficult to explain with the
metallicity dependent SN Ia model.

However, it is possible to produce such low [$\alpha$/Fe] with
core-collapse SNe.  As discussed above, yields from 13-15 \msun~ stars
have low (even sub-solar) [$\alpha$/Fe].  Thus if IMF in those dwarf
spheroidal galaxies is steep enough or the upper limit mass of
core-collapse SNe is truncated around 20 \msun~\cite{shet02}, the
contribution of 13-15 \msun~ stars could be significantly larger than
in our Galaxy.  Also it is possible that some fraction of AGB stars
suffers less mass loss because of low metallicity, thus growing
degenerate C+O cores to the Chandrasekhar mass before losing entire
envelopes.  If this is the case, those metal-poor AGB stars undergo SN
Ia-like explosions (SN I+1/2).  If the mass range is very narrow, say
$\sim$ 7.5-8.0 \msun, SNe I+1/2 do not affect [O/Fe] for our Galaxy if
integrated over the Salpeter IMF up to 50 \msun~\cite{nom91}.
However, if IMF is steeper or the upper-mass limit is lower,
contributions of SNe I+1/2 and 13-15 \msun~ stars are larger, thus
reducing [$\alpha$/Fe].

The abundance patterns in dSph as well as some Damped Lyman-$\alpha$
systems can provide important constraint on the progenitor mass ranges
of SNe Ia, SNe I+1/2, and SNe II.  Also the appearance of SNe Ia in
more metal-rich regions of dSph may not be impossible in view of spatial
inhomogeneity.

\section{The origin of diversity of type Ia supernovae}

 There are some observational indications that SNe Ia depend on the type
of the host galaxies.  The most luminous SNe Ia seem to occur only in
spiral galaxies, while both spiral and elliptical galaxies are hosts for
dimmer SNe Ia. Thus the mean peak brightness is dimmer in ellipticals
than in spiral galaxies \cite{ham96}.  Also the SNe Ia rate per unit
luminosity at the present epoch is almost twice as high in spirals as in
ellipticals \cite{cap97}.

 Umeda et al. \cite{ume99b} suggested that the variation of the C/O
ratio is the main cause of the variation of SNe Ia brightness, with
larger C/O ratio yielding brighter SNe Ia.  Here we will show that the
C/O ratio depends indeed on the metallicity and age of the companion of
the WD, and that the model can explain most of the observational trends
discussed above. We then make some predictions about the brightness of
SN Ia at higher redshift.

\subsection{C/O ratio in white dwarf progenitors}

 The C/O ratio in C+O WDs depends primarily on the main-sequence mass
of the WD progenitor and on metallicity.  The most important
metallicity effect is that the radiative opacity is smaller for lower
$Z$. Therefore, a star with lower $Z$ is brighter, thus having a
shorter lifetime than a star with the same mass but higher $Z$.  In
this sense, the effect of reducing metallicity for these stars is
almost equivalent to increasing a stellar mass.

 For stars with larger masses and/or smaller $Z$, the luminosity is
higher at the same evolutionary phase.  With a higher nuclear energy
generation rate, these stars have larger convective cores during H and
He burning, thus forming larger He and C-O cores.

\begin{figure}[t]
	\begin{center}
		\begin{minipage}[t]{0.8\textwidth}
		\includegraphics[width=0.9\textwidth]{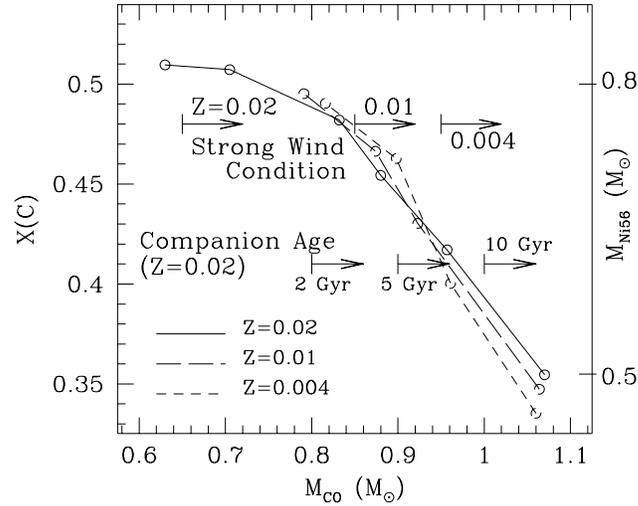}
		\end{minipage}
	\end{center}
\caption{
The total $^{12}$C mass fraction included in the convective core of
mass, $M=1.14M_\odot$, just before the SN Ia explosion as a function
of the C+O core mass before the onset of mass accretion, 
$M_{\rm CO}$. The lower bounds of $M_{\rm CO}$ obtained from the age
effects and the conditions for strong wind to blow are also shown by
arrows \cite{ume99b}.
\label{nomotoF2}}
\end{figure}

According to the evolutionary calculations for 3$-$9 $M_\odot$ stars
\cite{ume99a}, the C/O ratio and its distribution are determined in the
following evolutionary stages of the close binary.

(1) At the end of central He burning in the 3$-$9 $M_\odot$ primary
star, C/O$<1$ in the convective core. The mass of the core is larger
for more massive stars. 

(2) After central He exhaustion, the outer C+O layer grows via He
shell burning, where C/O$\gsim 1$ \cite{ume99a}.

(3a) If the primary star becomes a red giant (case C evolution), it then
undergoes the second dredge-up, forming a thin He layer, and enters the
AGB phase. The C+O core mass, $M_{\rm CO}$, at this phase is larger for
more massive stars. For a larger $M_{\rm CO}$ the total carbon mass
fraction is smaller.

(3b) When it enters the AGB phase, the star greatly expands and is
assumed here to undergo Roche lobe overflow (or a super-wind phase)
and to form a C+O WD. Thus the initial mass of the WD, $M_{\rm
WD,0}$, in the close binary at the beginning of mass accretion is
approximately equal to $M_{\rm CO}$.

(4a) If the primary star becomes a He star (case BB evolution), the
second dredge-up in (3a) corresponds to the expansion of the He
envelope.

(4b) The ensuing Roche lobe overflow again leads to a WD of $M_{\rm
WD,0}$ = $M_{\rm CO}$.

(5) After the onset of mass accretion, the WD mass grows through
steady H burning and weak He shell flashes, as described in the WD
wind model.  The composition of the growing C+O layer is assumed to be
C/O=1.

(6) The WD grows in mass and ignites carbon when its mass reaches
$M_{\rm Ia} =1.367 M_\odot$, as in the model C6 \cite{nom84}.
Carbon burning grows into a deflagration for a central temperature of
$8\times 10^8$ K and a central density of $1.47\times 10^9$ g
cm$^{-3}$.  At this stage, the convective core extends to $M_r =
1.14M_\odot$ and the material is mixed almost uniformly (model C6).

Figure \ref{nomotoF2} shows the carbon mass fraction $X$(C) in
the convective core of this pre-explosive WD, as a function of
metallicity ($Z$) and initial mass of the WD before the onset of mass
accretion, $M_{\rm CO}$.  We note: 

1) $X$(C) is smaller for larger $M_{\rm CO} \simeq M_{\rm WD,0}$.

2) The dependence of $X$(C) on metallicity is small when plotted
against $M_{\rm CO}$, even though the relation between $M_{\rm CO}$
and the initial stellar mass depends sensitively on $Z$ \cite{ume99a}.

\subsection{Brightness of type Ia supernovae and the C/O ratio}

In the Chandrasekhar mass models for SNe Ia, brightness of SNe Ia
is determined mainly by the mass of $^{56}$Ni synthesized ($M_{\rm
Ni56}$).  Observational data suggest that $M_{\rm Ni56}$ for most SNe
Ia lies in the range $M_{\rm Ni56} \sim 0.4 - 0.8 M_\odot$
(e.g., \cite{maz98}). 

Here we postulate that $M_{\rm Ni56}$ and consequently brightness of a
SN Ia increase as the progenitors' C/O ratio increases (and thus
$M_{\rm WD,0}$ decreases).  As illustrated in Figure \ref{nomotoF2},
the range of $M_{\rm Ni56} \sim 0.5-0.8 M_\odot$ is the result of an
$X$(C) range $0.35-0.5$, which is the range of $X$(C) values of our
progenitor models.  The $X$(C) -- $M_{\rm Ni56}$ -- $M_{\rm WD,0}$
relation we adopt is still only a working hypothesis, which needs to
be proved from studies of the turbulent flame during explosion
(e.g., \cite{hil00}).

\subsection{Metallicity and age effects}

Assuming the relation between $M_{\rm Ni56}$ and $ X$(C) given in
Figure \ref{nomotoF2}, the model predicts the absence of brighter SNe
Ia in lower metallicity environment.

In this model, the age of the progenitor system also constrains the
range of $X$(C) in SNe Ia. In the SD scenario, the lifetime of the
binary system is essentially the main-sequence lifetime of the companion
star, which depends on its initial mass $M_2$.  In order for the WD mass
to reach $M_{\rm Ia}$, the donor star should transfer enough material at
the appropriate accretion rates.  The donors of successful SN Ia cases
are divided into two categories: one is composed of slightly evolved
main-sequence stars with $M_2 \sim 1.7 - 3.6M_\odot$ (for $Z$=0.02), and
the other of red-giant stars with $M_2 \sim 0.8 - 3.1M_\odot$ (for
$Z$=0.02) \cite{kob02}.

If the progenitor system is older than 2 Gyr, it should be a system
with a donor star of $M_2 < 1.7 M_\odot$ in the red-giant branch.
Systems with $M_2 > 1.7 M_\odot$ become SNe Ia in a time shorter than
2 Gyr.  Likewise, for a given age of the progenitor system, $M_2$ must
be smaller than a limiting mass. This constraint on $M_2$ can be
translated into the presence of a minimum $M_{\rm CO}$ for a given
age, as follows: For a smaller $M_2$, i.e. for the older system, the
total mass which can be transferred from the donor to the WD is
smaller. In order for $M_{\rm WD}$ to reach $M_{\rm Ia}$, therefore,
the initial mass of the WD, $M_{\rm WD,0} \simeq M_{\rm CO}$, should
be larger.  This implies that the older system should have larger
minimum $M_{\rm CO}$ as indicated in Figure \ref{nomotoF2}.  Using the
$X$(C)-$M_{\rm CO}$ and $M_{\rm Ni56}$-$X$(C) relations
(Fig. \ref{nomotoF2}), we conclude that WDs in older progenitor systems
have a smaller $X$(C), and thus produce dimmer SNe Ia.

\subsection{Morphology of the host galaxies}

 Among the observational indications which can be compared with our
model is the possible dependence of the SN brightness on the
morphology of the host galaxies.  Hamuy et al. \cite{ham96} found that
the most luminous SNe Ia occur in spiral galaxies, while both spiral
and elliptical galaxies are hosts to dimmer SNe Ia. Hence, the mean
peak brightness is lower in elliptical than in spiral galaxies.

 In our model, this property is simply understood as the effect of the
different age of the companion. In spiral galaxies, star formation
occurs continuously up to the present time. Hence, both WD+MS and
WD+RG systems can produce SNe Ia. In elliptical galaxies, on the other
hand, star formation has long ended, typically more than 10 Gyr
ago. Hence, WD+MS systems can no longer produce SNe Ia. 
Since a WD with smaller $M_{\rm CO}$ is assumed to produce a brighter
SN Ia (larger $M_{\rm Ni 56}$), our model predicts that dimmer SNe Ia
occur both in spirals and in ellipticals, while brighter ones occur
only in spirals.  The mean brightness is smaller for ellipticals and
the total SN Ia rate per unit luminosity is larger in spirals than in
ellipticals.  These properties are consistent with observations.

\subsection{Evolution of type Ia supernovae at high redshift}

\begin{figure}[t]
	\begin{minipage}[t]{1.\textwidth}
		\includegraphics[width=1.\textwidth]{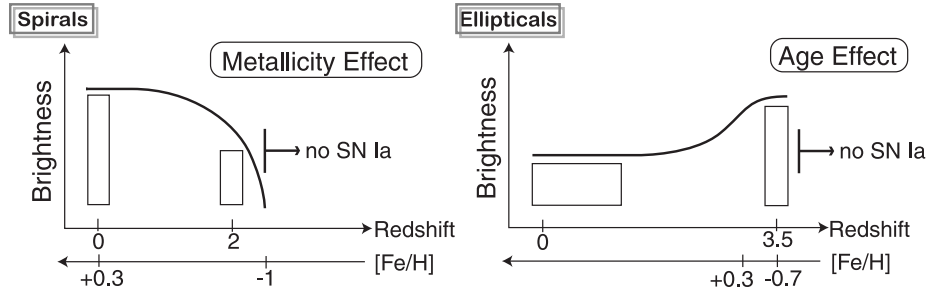}
	\end{minipage}
\caption{
Illustration of the predicted variation in SN Ia brightness with
redshift.
\label{nomotoF3}}
\end{figure}

 Our model predicts that when the progenitors belong to an old
population, or to a low metal environment, the number of very bright
SNe Ia is small, so that the variation in brightness is also smaller,
which is shown in Figure \ref{nomotoF3}. In spiral galaxies, the
metallicity is significantly smaller at redshifts $z\gsim 1$, and thus
both the mean brightness of SNe Ia and its range tend to be smaller
(Fig. \ref{nomotoF3}).  At $z\gsim 2$ SNe Ia would not occur in
spirals at all because the metallicity is too low.  In elliptical
galaxies, on the other hand, the metallicity at redshifts $z \sim 1-3$
is not very different from the present value.  However, the age of the
galaxies at $z\simeq 1$ is only about 5 Gyr, so that the mean
brightness of SNe Ia and its range tend to be larger at $z\gsim 1$
than in the present ellipticals because of the age effect.

 We note that the variation of $X$(C) is larger in metal-rich nearby
spirals than in high redshift galaxies.  Therefore, if $X$(C) is the
main parameter responsible for the diversity of SNe Ia, and if the light
curve shape (LCS) method is confirmed by the nearby SNe Ia data
\cite{rie95,ham95}, the LCS method can also be used to determine the
absolute magnitude of high redshift SNe Ia.

Here we consider the metallicity effects only on the C/O ratio; this
is just to shift the main-sequence mass - $M_{\rm WD,0}$ relation,
thus resulting in no important evolutionary effect.  However, some
other metallicity effects could give rise to evolution of SNe Ia
between high and low redshifts (i.e., between low and high
metallicities) (see, e.g., \cite{nom00}).

\section{Rotation}

The accreting WD gains angular momentum from the rotating disk in
addition to the mass, thus rotating faster and faster.  Rapid rotation
of a WD affects its limiting mass and final structure.  The effects of
rotation on the evolution of accreting WDs have been studied mostly
with 1D approximate models (e.g., \cite{yoon02,pier02,saio02}).

\subsection{Diversity and rotation}

\begin{figure}[t]
	\begin{center}
	\begin{minipage}[t]{0.8\textwidth}
		\includegraphics[width=0.9\textwidth]{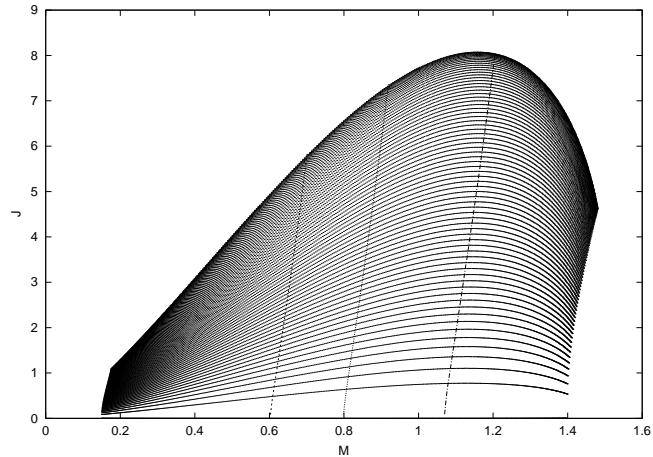}
	\end{minipage}
	\end{center}
\caption{
The evolutionary track in the total angular momentum $J$ (10$^{49}$
erg sec) and the WD mass $M$ (\msun) of uniformly rotating WDs
starting from the initial masses of 0.6 \msun, 0.8 \msun, and 1.07
\msun~ \cite{ueni02}.
\label{nomotoF4}}
\end{figure}

Uenishi et al. \cite{ueni02} have calculated the axisymmetric
structure of {\sl uniformly} rotating WDs and followed the
evolutionary sequence in the total angular momentum $J$ and the WD
mass $M$ (Fig. \ref{nomotoF4}).  Here accreting gas is supposed to obey
Keplerian rotation law and gives its angular momentum to the WD.  The
almost straight lines show the evolutionary tracks of the accreting
WDs starting from the initial masses of 0.6 \msun, 0.8 \msun, and 1.07
\msun, respectively.  After gaining $\sim$ 0.1 \msun, the WDs reach
the critical rotation at the upper edge of the $J-M$ diagram where
the equatorial centrifugal force is equal to the gravity.

For WDs which have reached the critical rotation, Paczy\'nski
\cite{pac90} and Pooham \& Narayan \cite{pop90} found a solution that
permits the WD to accrete without becoming secularly unstable.  In such
a solution, the angular momentum is transported backwards from the WD to
the disk while the mass of the WD increases. Then the WD evolves along
the upper envelope in the $J$-$M$ plane and eventually explodes at the
upper-right corner.  (Along the right edge of the diagram, the central
density of WDs is 2 $\times$ 10$^9$ g cm$^{-3}$ where carbon is
ignited.)  Then all the rapidly rotating WDs explode with the same
$(M,J)$.  In other words, {\sl uniform} rotation leads to the uniformity
rather than the diversity of SNe Ia.

However, rotation could produce the diversity of SNe Ia in the following
way.  The uniform rotation is realized if the timescale of angular
momentum transport in the WDs is much shorter than the accretion time
scale.  If the angular momentum transport is much slower, then the
initially non-rotating part of the WD may remain non-rotating, while the
accreted outer part of the WD rotates fast and reaches the critical
rotation.  When the WD reaches the carbon ignition at the limiting mass,
the angular momentum brought into the WD is larger (smaller) for smaller
(larger) $M_0$.  The WD with smaller $M_0$ grows to larger
$M_{\mathrm{fin}}$ at the explosion.  Although $M_{\mathrm{fin}}$ $\sim$
1.45 \msun~ are not so different from each other, larger
$M_{\mathrm{fin}}$ could produce brighter SNe Ia.

Also the distribution of the angular velocity is different, i.e., the
non-rotating core is smaller (larger) for smaller (larger) $M_0$.
Such a difference might affect the flame propagation and lead to 
different amount of $^{56}$Ni.

If the WD with smaller $M_0$ produce a brighter SN Ia with the effect of
rotation, the upper limit of SNe Ia brightness is higher in spirals than
in ellipticals.  SNe Ia in spirals can originate from WDs with a wide
range of $M_0$, thus having a large dispersion of brightness than those
in ellipticals. This could explain the observation that the most
luminous SNe Ia appear to be observed only in spirals, while dimmer SNe
Ia are observed in both spirals and ellipticals.

\subsection{Rotation and rapid accretion in merging white dwarfs}

Nomoto \& Iben \cite{nom85} and Saio \& Nomoto \cite{saio98} have
simulated the merging of double WDs in 1D and shown that the rapidly
accreting WDs undergo off-center carbon ignition if \mdot~ $\gsim$ 2
$\times$ 10$^{-6}$ \msy~ because of compressional heating.  Afterwards
carbon flame propagates inward through the center and converts C+O into
O+Ne+Mg.  Then the final outcome is most likely accretion-induced
collapse rather than SNe Ia.

Recently Piersanti et al. \cite{pier02} calculated the evolution of
WDs with rotation in 1D approximation and argued that the lifting
effect of rotation reduces compressional heating, and the WD could
avoid off-center carbon ignition and reach the central carbon ignition
to produce SNe Ia.  However, they did not follow the accretion after
the critical rotation is reached because the backward transport of
angular momentum to the disk was not taken into account.

Saio \& Nomoto \cite{saio02} have also calculated the accretion of C+O
onto the C+O WD with rotation for various timescale of angular
momentum transport in 1D approximation.  The outermost layer of the
accreting WD quickly reaches the critical rotation as in Figure
\ref{nomotoF4}.  Afterwards, the angular momentum is transported
backward to disk and accretion continues.  For \mdot $\gsim$ 4
$\times$ 10$^{-6}$ \msy, off-center carbon burning is ignited prior to
the central C-ignition.  Thus the lifting effect of rotation increases
the critical accretion rate for the occurrence of off-center
C-ignition by a factor of $\sim$ 2 compared with the non-rotating
case, but the basic conclusion is the same as non-rotating case, i.e.,
the accretion-induced collapse is the most likely outcome in the DD
scenario \cite{saio98}.

\section {SN 2000cx}

\begin{figure}[t]
	\begin{center}
	\begin{minipage}[t]{0.85\textwidth}
		\includegraphics[width=0.9\textwidth]{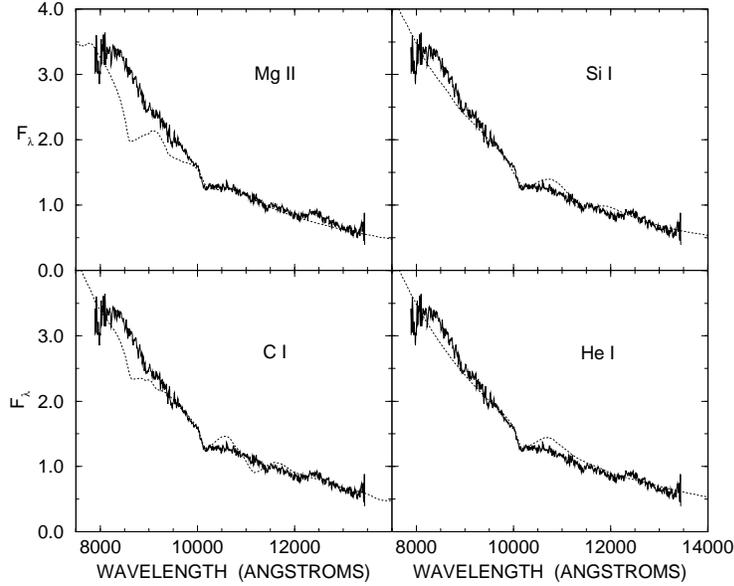}
	\end{minipage}
	\end{center}
\caption{ The average of two infrared spectra of SN~2000cx 
obtained by Rudy et~al.\cite{rudy02} at the Lick Observatory on 2000 July
band (solid lines) is compared with four synthetic spectra (dotted
lines), each of which contain lines of only one ion \cite{hata02}.
\label{nomotoF5}}
\end{figure}

Supernova 2000cx was a very well observed SN~Ia that in certain respects
resembled the peculiar ``powerful'' SN~1991T, but also showed some
photometric and spectroscopic characteristics that are unprecedented
among well observed SNe~Ia (Li et al. \cite{li01}; hereafter L01).

\subsection {Infrared spectra}

Rudy et~al. (\cite{rudy02}; hereafter R02) obtained infrared spectra
covering the range 0.8 to 2.5~$\mu$m, 6 and 5 days before optical
maximum.  They attributed an absorption feature near 1.0~$\mu$m to
Mg~II \lam10926.  Hatano et al. \cite{hata02} have used spectrum code
SYNOW and found that the absorption feature is most likely due to
He~I \lam10830.  The presence of helium would have important
implications for models.

In Figure \ref{nomotoF5} the infrared spectrum \cite{rudy02} is compared
with four synthetic spectra, each containing lines of only one ion.  The
synthetic spectra have $T_{\rm bb} = T_{\rm exc} = 12,000$~K and $v_{\rm
phot} = 22,000$ \kms\ (Mg II), 16,000 \kms\ (C~I), and 20,000 \kms\
(Si~I and He~I).

Mg~II \lam10926 has been considered as a possible identification for the
absorption feature by R02.  However, the upper left panel of Figure
\ref{nomotoF5} shows that when \lam10926 is strong enough to account for
the 1.0~$\mu$m absorption, other Mg~II lines produce unwanted
absorptions near 8600~\AA\ (\lam9226) and 9400~\AA\ (\lam9632).  In LTE
\lam9226 has a larger optical depth than \lam10926 for any reasonable
excitation temperature.  It is unlikely that Mg~II is responsible for
the 1.0~$\mu$m feature.

The lower left panel of Figure \ref{nomotoF5} shows that C~I, with
\lam10695 producing the 1.0~$\mu$m absorption, has a similar problem of
unwanted features, due to \lam9055 and \lam11755.  The upper right panel
shows that a group of Si~I lines (mainly a multiplet at \lam10790) fit
the 1.0~$\mu$m feature without severe problems; a feature due to
\lam12047 is from the same lower level and is stronger, but one might
argue that it is possibly present, blended with other lines, in the
observed spectrum.  However, unless the level of ionization in SN~2000cx
was much lower at the epoch of the infrared spectrum than it was at the
time of the first optical spectrum, just a few days later, Si~I lines
would not be expected.  Finally, the lower right panel of Figure
\ref{nomotoF5} shows that He~I \lam10830, with a optical depth at the
photosphere of 0.9, can account for the 1.0~$\mu$m feature without
causing any obvious problems.  By this process of elimination, Hatano et
al. \cite{hata02} find that the most likely identification of the
1.0~$\mu$m feature in SN~2000cx is He~I $\lambda$10830.  The evidence
for helium in SN~2000cx, although not conclusive, is strong enough to be
taken seriously.

\subsection {Sub-Chandrasekhar mass model}

Both L01 and R02 suggest a delayed detonation as a favorable model for
SN 2000cx.  In particular, R02 argued that the 1.0~$\mu$m feature is
due to Mg II and the presence of such high velocity Mg as $\gsim$
20,000 \kms\ supports the delayed detonation model.  However, this
argument does not hold if the 1.0~$\mu$m feature is due to He I.
Rather the delayed detonation models do not contain such high velocity
He (e.g., \cite{iwa99}).

The presence of He, instead, may suggest the sub-Chandrasekhar model for
SN 2000cx.  In the sub-Chandrasekhar mass models, the He detonation in
the outer He layer synthesize mostly $^{56}$Ni but significant amount of
He is also present due to strong $\alpha$-rich freezeout
\cite{nom82b,livne95,ww94,nom97}.  The velocities of He are $\sim$
11,000 - 14,000 \kms\ near the bottom of the He layer and $\sim$ 25,000 -
30,000 \kms\ near the surface, depending on $M_{\rm WD}$ and the mass of
the He layer ($M_{\rm He}$).  The He velocities and the $^{56}$Ni mass
of SN 2000cx could be consistent with the model with relatively large
$M_{\rm WD}$ and $M_{\rm He}$.  In the outer detonated layers, little Si
is produced, which is also consistent with the weak Si features of SN
2000cx.  

The presence of high velocity He with $^{56}$Ni (and little Si) is also
seen in the late detonation model W7DHE of the Chandrasekhar mass white
dwarf \cite{yam92}.  The explosion of this type would occur for slower
accretion than the sub-Chandrasekhar mass explosion \cite{nom82a}.

%

\end{document}